\documentclass[conference,letterpaper]{IEEEtran}

\usepackage[utf8]{inputenc} 
\usepackage[T1]{fontenc}
\usepackage{url}
\usepackage{ifthen}
\usepackage{cite}
\usepackage[cmex10]{amsmath} 

\usepackage{algorithmic, algorithm}
\usepackage[english]{babel}
\usepackage{amssymb}
\usepackage{amsmath}
\usepackage[dvipsnames]{xcolor}
\usepackage{cite}
\usepackage{graphicx}
\usepackage{balance}
\usepackage{mathtools}
\usepackage{comment}
\usepackage[a4paper, total={6.9in, 9.2in}]{geometry}

\newtheorem{theorem}{Theorem}
\newtheorem{lemma}{Lemma}

\newtheorem{construction}{Construction}

\newtheorem{definition}{Definition}

\addto\extrasenglish{

}

\begin{document}
\title{A Single-Bit Redundancy Framework for Multi-Dimensional Parametric Constraints} 

\IEEEoverridecommandlockouts

\author{%
  \IEEEauthorblockN{Daniella~Bar-Lev}
  \IEEEauthorblockA{Center for Memory and Recording Research\\
                    University of California San Diego, CA, USA\\
                    Email: dbarlev@ucsd.edu}
  \and
  \IEEEauthorblockN{Michael Shlizerman}
  \IEEEauthorblockA{The Department of Physics\\ 
                    Technion – Israel Institute of Technology, Haifa, Israel\\
                    Email: mshlizerman@campus.technion.ac.il}
\thanks{The work of D. Bar-Lev was supported in part by Schmidt Sciences and by NSF Grant CCF2212437.}}

\maketitle

\begin{abstract} 
    Constrained coding plays a key role in optimizing performance and mitigating errors in applications such as storage and communication, where specific constraints on codewords are required. While non-parametric constraints have been well-studied, parametric constraints, which depend on sequence length, have traditionally been tackled with ad hoc solutions. Recent advances have introduced unified methods for parametric constrained coding. This paper extends these approaches to multidimensional settings, generalizing an iterative framework to efficiently encode arrays subject to parametric constraints. We demonstrate the application of the method to existing and new constraints, highlighting its versatility and potential for advanced storage systems.
\end{abstract}

\section{Introduction}
Constrained coding plays a fundamental role in research on information theory and coding theory, focusing on the design of codes that adhere to specific constraints imposed on the codewords. These constraints are often imposed to enhance system performance or prevent operational failures, as seen in applications ranging from optical and magnetic storage to modern innovations like DNA-based data storage (see e.g.,~\cite{immink2004codes,marcus2001introduction, immink2022innovation}). 

For example, run-length limited (RLL) constraints~\cite{marcus2001introduction,van2010construction,levy2018mutually, immink1990runlength} in magnetic storage mitigate read errors caused by long sequences of identical bits, while balanced codes are vital in optical communication systems to reduce baseline wander and in DNA-based storage system to reduce media  degradation~\cite{knuth, TB99, IW08, WISS13, nguyen2020binary}. These practical demands have driven extensive research into designing efficient constrained codes that maximize the rate while ensuring that all outputs comply with the imposed restrictions.

Constrained coding problems can be broadly categorized along two axes: \emph{dimensionality} (one-dimensional or multidimensional constraints) and \emph{parametricity} (parametric or non-parametric constraints). Non-parametric constraints are fixed and independent of the sequence's size. A classic example is the $d$-zero-block-avoiding constraint, where the sequence cannot have $d$ or more consecutive zeros (see e.g.,~\cite{marcus2001introduction}), a constraint determined solely by the parameter $d$ and not the length of the sequence. Parametric constraints, in contrast, depend on the length of the encoded sequence. For instance, a parametric zero-block avoidance constraint might forbid blocks of zeros whose length increases with the sequence size~\cite{levy2018mutually, van2010construction, kobovich2024universal, bar2023universal} (e.g., no more than $\log(n)$ consecutive zeros, where $n$ is the codeword length).

Non-parametric constraints in one-dimensional settings have been extensively explored, and a well-established framework exists for designing solutions that handle a wide range of such constraints~\cite{marcus2001introduction}. The general approach begins with constructing a deterministic finite automaton (DFA) that recognizes all valid words under the given constraint. This is followed by the application of the state-splitting algorithm~\cite{adler1983algorithms}, which facilitates the construction of efficient encoders that approach the channel capacity. Furthermore, channel capacity can be determined using the Perron-Frobenius theorem~\cite{Min88}, which links capacity to the spectral radius of the adjacency matrix of the DFA.

In contrast, parametric constraints were traditionally tackled with specific, ad hoc solutions tailored to individual cases. However, recent progress has led to the development of unified methods for parametric constrained coding. Notably, in~\cite{bar2023universal, kobovich2024universal} an iterative algorithm with guaranteed convergence and efficient average time complexity was introduced. This technique was also utilized in~\cite{bar2024balance, ConstrainedPeriodicity} to tackle specific parametric constraints. Additionally, Ryabko~\cite{ryabko2024general} developed a method based on enumerative coding. These advances represent significant progress in the systematic treatment of parametric constraints.

Although constrained coding in one dimension has been extensively studied, multidimensional constrained coding introduces unique challenges and opportunities. In two-dimensional (2D) arrays, constraints can model practical requirements in storage devices such as flash memory or data grids for DNA storage. Notable works include, for example,~\cite{sharov2010two, tal2009row, talyansky1999efficient, ordentlich2012asymptotic, ordentlich2012low},
however, most of these studies focus on non-parametric constraints. The work by Marcovich and Yaakobi~\cite{marcovich2023zero} is one of the few that addresses parametric multidimensional constraints, studying forbidden sub-arrays whose size scales with the dimensions of the array.

In this work, we extend the universal approach for parametric constrained coding introduced in \cite{bar2023universal, kobovich2024universal} to multidimensional settings. Our method generalizes the iterative framework to efficiently encode arrays that satisfy a range of parametric constraints, including the zero-block avoidance constraint and the repeat-free constraints studied in~\cite{marcovich2023zero}. Moreover, we demonstrate the application of our approach to additional, previously unsolved constraint, further highlighting its versatility. 


\section{Definitions}\label{sec:background}
In this section, we formally define the notations that are used in this paper. Let $\Sigma\triangleq \{0,1\}$ be the binary alphabet, and for integers $n>0$ and $d>0$, we let $$n^d\triangleq \underbrace{n\times n\times \ldots \times n}_{d \text{ times}},$$
and $\Sigma^{n^d}$ is the set of all binary $d$-dimensional arrays of size~$n^d$. A position in an array $A\in\Sigma^{n^d}$ is a $d$-tuple ${\mathbf{I} = (i_1,i_2,\ldots, i_d)}$, such that $0\le i_j\le n-1$ for $1\le j\le  d$.

\begin{definition}
    (Sub arrays). Given $A\in\Sigma^{n^d}$, a sub-array starting at position $\mathbf{I}=(i_{1},i_{2},\dots i_{d})$ with size $\mathbf{d=(}\ell_{1},\ell_{2},\dots \ell_{d})$, for $1\le \ell_j\le n-i_j$ is denoted by $A_{\mathbf{I},\mathbf{d}}$. We denote by $R\left(A_{\mathbf{I},\mathbf{d}} \right)$ the part of $A$ that remains after erasing $A_{\mathbf{I},\mathbf{d}}$.
\end{definition}

    \begin{figure}[h]
    \centering \includegraphics[width=.8\linewidth]{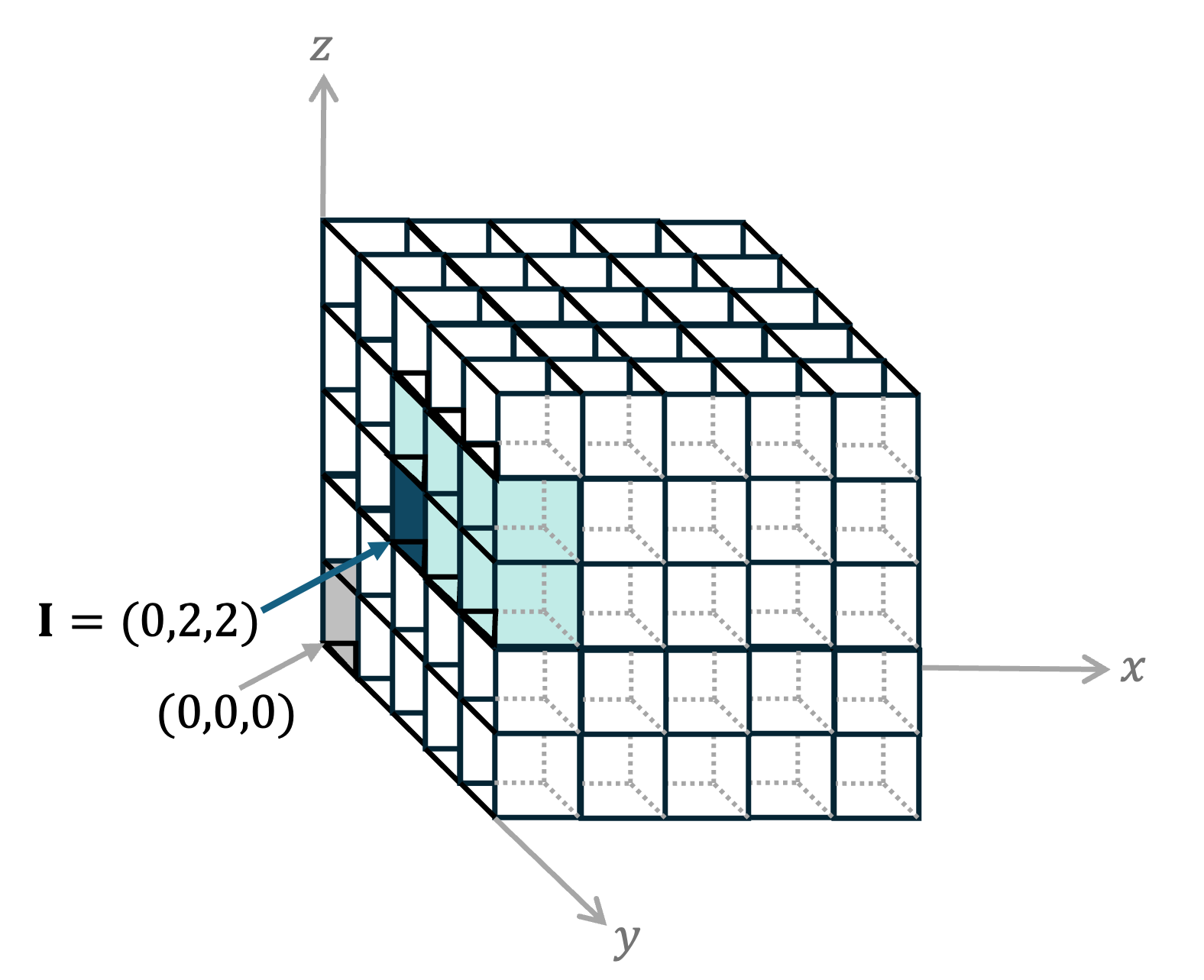}
    \caption{Illustration of an array $A\in\Sigma^{5^3}$ (i.e., $n=5$ and $d=3$), with a sub-array $A_{\mathbf{I},\mathbf{d}}$, starting at position $\mathbf{I} = (0,2,2)$ and size $\mathbf{d}=(1,3,2)$. The starting position $\mathbf{I}$ is highlighted with dark blue and the rest of the sub-array is highlighted with light blue.}
    \label{fig:const_min}
\end{figure}

Next, we extend the definition of parametric constraints presented in~\cite{bar2023universal,kobovich2024universal} to the multi-dimensional setting. 
Parametric constraints are conditions enforced on arrays (or strings) where the constraint itself
varies according to the message length. 
 For instance, in a string of length
$n$, there should be no two identical substrings of length
$\ell(n)$, where $\ell(n)$ is a function of $n$. This example can be extended to the $d$-dimensional setting where an array of size $n^d$ should not contain two identical sub-arrays of size $\ell(n)$.  A formal definition of such constraints is provided below. 
\begin{definition}
(Parametric Constraint). A parametric constraint $\mathcal{C}$ applied to a $d$-dimensional array channel of size $n^d$, means that a message $A\in\Sigma^{n^d}$ is accepted only if $A\in \mathcal{C}\left(n,d\right) \triangleq \mathcal{C}\cap \Sigma^{n^d}$.
\end{definition}

\begin{definition}
    (Constrained Encoder and Decoder). A parametric channel encoder $f_{k}:\Sigma^{k}\to \mathcal{C}\left(n,d\right)$ encodes general messages of length $k$ to an array of size $n^{d}=k+r$ that satisfy the constraint $\mathcal{C}\left(n,d\right)$, where $r$ is the redundancy of the encoder.
    Given an encoder $f_{k}$, a decoder is a function $g_{k}:\mathcal{C}\left(n, d\right)\to\Sigma^{k}$
that satisfies $g_{k}\left(f_{k}\left(x\right)\right)=x$ for all $x\in\Sigma^{k}$.
\end{definition}

In this work, we focus on the case of a single redundancy bit, that is, $r=1$. 
Note that although we focus on the binary alphabet  $\Sigma = \{0,1\}$,  all the definitions and results can be generalized to any alphabet size. 

\subsection*{Sub-Array Deletion}
The universal approach presented in~\cite{bar2023universal, kobovich2024universal} is based on an iterative algorithm in which, in each iteration, the validity of the string is tested using an indicator function, and in case the string is invalid, an injective function is applied to convert the forbidden sequence to another sequence. This process is repeated until the new sequence satisfies the constraint. In the special case where the constraint is \emph{local}, i.e., there is a set of forbidden substrings (of length $\ell(n)$) that should not appear in the encoded message, each iteration of this process can be performed as follows.
\begin{itemize}
    \item To test if the string is valid, each substring of length $\ell(n)$ is compared against the set of forbidden substrings. 
    \item In case a forbidden substring is found, this substring is deleted from the string and a compressed version of it,  together with its location and an additional bit, are appended to the end of the string. 
\end{itemize}
For more details, see~\cite[Construction 4]{bar2023universal}. Extending the validity verification to the multi-dimensional setup is straightforward. Next, we explain the procedure of sub-array deletion, which is required for the second step, and will be used throughout this paper. 

\begin{figure}[h]
    \centering \includegraphics[width=.995\linewidth]{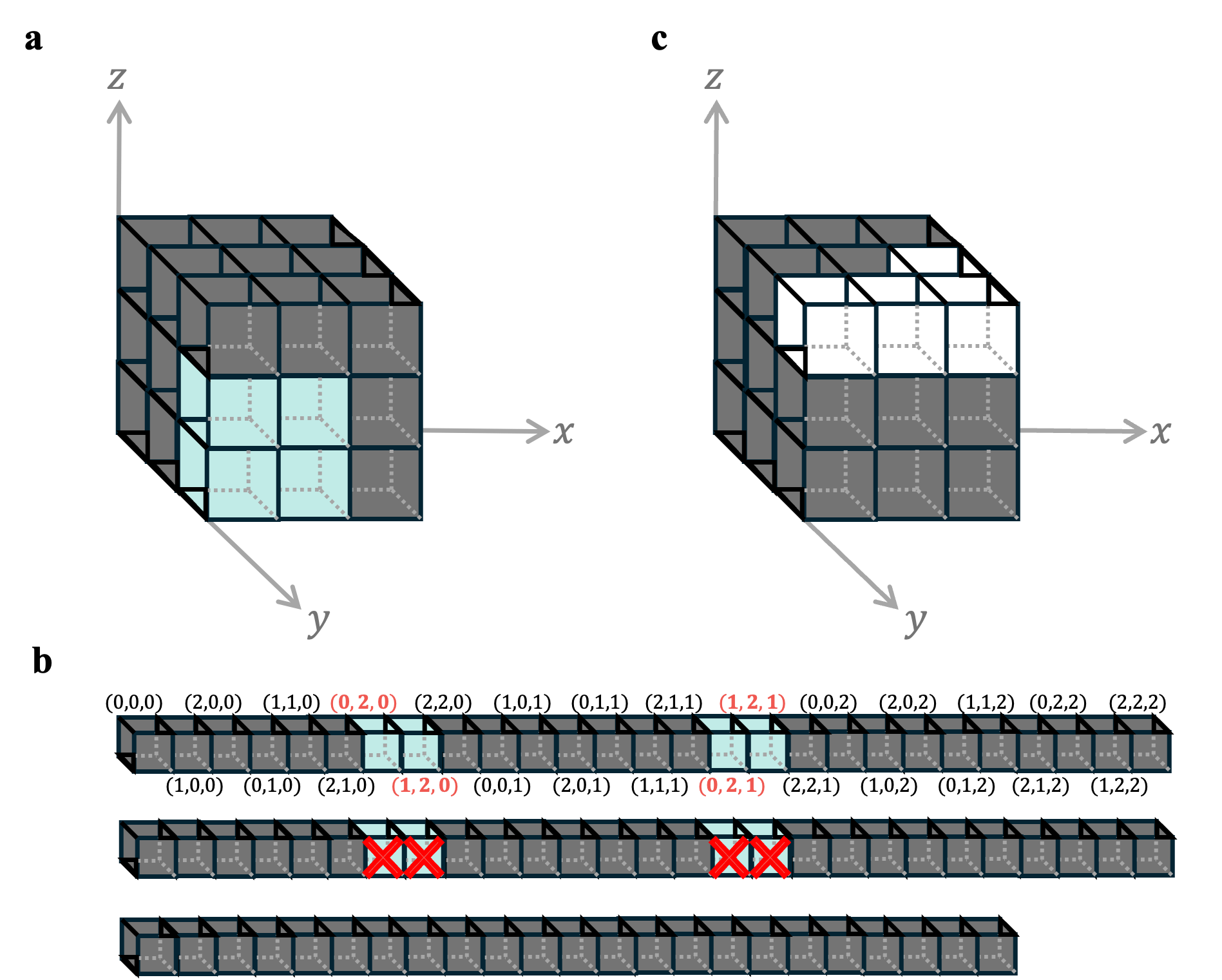}
    \caption{
Illustration of the vectorization process for an array $A \in \Sigma^{3^3}$ into a one-dimensional array, the deletion of a sub-array, and reconstruction into an almost-complete array. 
(a) The original array $A$ with the sub-array to be deleted highlighted in blue. 
(b) The vectorized representation of $A$, showing the erasure of the blue sub-array. 
(c) The almost-complete array reconstructed from the one-dimensional vector following the deletion process.
}
    \label{fig: delete sub-array}
\end{figure}

To remove a sub-array $A_{\mathbf{I},\mathbf{d}}$ from a given array \( A \), the following process can be used. First, the array $A$ is vectorized to a one-dimensional array, such that the element in location ${\mathbf{I} = (i_1,i_2,\ldots, i_d)}$ in $A$ appears in index \[
i_1 + i_2 \cdot n + \dots +i_{d-1}\cdot n^{d-2} + i_d \cdot n^{d-1}
\]
of the one-dimensional string. Then, all the symbols that correspond to the sub-array $A_{\mathbf{I},\mathbf{d}}$ are deleted from the obtained string. Finally, the remaining symbols are organized again in an array of size $n^d$, where the entries that correspond with the ``last" elements remain empty. The array obtained by this process is denoted by $R(A_{\mathbf{I},\mathbf{d}})$. This process is illustrated in Figure~\ref{fig: delete sub-array}.  

The time complexity of this process is linear with respect to the number of symbols in the array, i.e., $\mathcal{O}(n^d)$. This is because the process involves two iterations over the array: the first to construct the one-dimensional string and the second to reconstruct the almost full $d$-dimensional array.  

In the discussion section, we briefly describe a more efficient method for deleting sub-arrays; however, a detailed description of this approach is left for the long version of the paper.

\section{Universal Framework for Multi-Dimensional Constrained Codes}
In this section, we extend the construction of universal encoder and decoder algorithms shown in \cite[Construction 2]{bar2023universal} to the multi-dimensional constrained channels. To this end, we use the notation $\Sigma^{n^d-1}$ to denote the set of arrays of size $n^d$ in which the entry at location ${\mathbf{I}_{n-1}=(n-1,n-1,\ldots,n-1)}$ is empty. Furthermore, for a constraint $\mathcal{C}(n,d)\subseteq \Sigma^{n^d}$, we denote the set of invalid arrays by $\overline{\mathcal{C}(n,d)}\triangleq \Sigma^{n^d}\setminus \mathcal{C}(n,d)$.  

\begin{theorem}\label{th: construction existence}
    Given a parametric constraint $\mathcal{C}(n,d)\subseteq \Sigma^{n^d}$, 
    there exists an encoder-decoder pair for  $\mathcal{C}(n,d)$ with a single redundancy bit if the following exist:
    \begin{enumerate}
    \item An indicator $\mathbf{1}_{\mathcal{C}(n,d)}:\Sigma^{n^d}\rightarrow\Sigma$  for the set $\mathcal{C}(n,d)$. 
    \item An injective function $\xi: \overline{\mathcal{C}(n,d)}\rightarrow\Sigma^{n^d-1}$.
\end{enumerate}
\end{theorem}

The proof of Theorem~\ref{th: construction existence} is based on presenting the encoder and decoder algorithms and proving their correctness. As both the algorithms and the proof of their correctness are very similar to the proof presented in~\cite{bar2023universal}, we describe the encoder and decoder for the multi-dimensional case for completeness, and refer the reader to~\cite{bar2023universal} for the proof of their correctness (and, in particular, the converges of the encoder).  

\textbf{Encoder function $f$.} Given an input string $x$ of $2^{n^d-1}$~bits, we first embed it into an array $A' \in \Sigma^{n^d-1}$ (similarly to the last step in the sub-array deletion discussed in Section~\ref{sec:background}) and fill the empty entry with $0$ to obtain an array $A\in\Sigma^{n^d}$. 
Subsequently, while $\mathbf{1}_{\mathcal{C}(n,d)}(A)=0$, i.e., $A\notin \mathcal{C}(n,d)$, let $A'\leftarrow \xi(A)$ and update $A$ to be $A'$ with a $1$ in the empty entry. 
This iterative process continues until $\mathbf{1}_{\mathcal{C}(n,d)}(A)=1$, and the obtained $A$ is the output of the encoder.

\textbf{Decoder function $g$.} Given an array \( A \in \Sigma^{n^d} \), while the value at location $\mathbf{I}_{n-1}=(n-1,\ldots,n-1)$ is equal to $1$, delete the symbol at location $\mathbf{I}_{n-1}$ to obtain an array $A'\in\Sigma^{n^d-1}$ and update $A$ with $\xi^{-1}(A')$. 
Once the while loop terminates (i.e., the symbol at location $\mathbf{I}_{n-1}$ is $0$), the decoder returns the remaining $n^d - 1$ bits as a string.

We note that by the same arguments presented in~\cite{bar2023universal}, the average time complexity of $f$ and $g$ is $\mathcal{O}( T(n^d))$, where $T(n^d)$ is the maximal time complexity among $\xi, \xi^{-1},\mathbf{1}_{C(n,d)}$.

\section{Constrained Sub-Arrays}

\subsection{Zero-Rectangular-Cuboid-Free (ZRCF)}
    The Zero-Rectangular-Cuboid-Free (ZRCF) constraint prohibits the appearance of an all-zero rectangular cuboid of a given size as a sub-array. More formally, for integers $n,d$ and a given size $\mathbf{d} = (\ell_1,\ldots, \ell_d)$, the $\mathbf{d}$-ZRCF constraint is defined as follows, 
    \[
    \mathcal{C}_{\mathbf{d}-ZRCF}\left(n, d\right) = \left\{ A\in\Sigma^{n^{d}}:
    \forall \mathbf{I},  A_{\mathbf{I}, \mathbf{d}}\ne \overline{\mathbf{0}_{\mathbf{d}}} \right\}, 
    \]
    where $\overline{\mathbf{0}_{\mathbf{d}}}$ is the all-zero rectangular cuboid of size $\mathbf{d}$, and we only consider locations $\mathbf{I}$ for which $A_{\mathbf{I}, \mathbf{d}}$ is well defined. 
    \begin{construction}\label{const: ZRCF}
    We define $\mathbf{1}_{\mathcal{C}(n,d)}:\Sigma^{n^d}\rightarrow\Sigma$ to be a function that scans all the rectangular cuboid of size $\mathbf{d}$ in the input array $A$ and returns $1$ if and only if none of them is the all-zero rectangular cuboid. Additionally, we define an injective function
    $
    \xi:\overline{C_{\mathbf{d}-ZRCF}\left(n, d\right)}\rightarrow\Sigma^{n^{d}-1}
    $
    such that for $A\in\overline{C_{\mathbf{d}-ZRCF}\left(n, d\right)}$,
    \[\xi\left(A\right)=R\left(A_{\mathbf{I},\mathbf{d}}\right)\circ \mathbf{I},
    \]
    where  $A_{\mathbf{I},\mathbf{d}}$
    is the first\footnote{Here, first refers to a standard per-coordinate ordering} all-zero rectangular cuboid of size $\mathbf{d}$, and $\circ$ refers to filling the binary representation of $\mathbf{I}$ in the (all except the last) empty entries of $\left(A_{\mathbf{I},\mathbf{d}}\right)$. Note that such sub-array exists as $A\notin{C_{\mathbf{d}-ZRCF}\left(n, d\right)}$.
    
    As we need a sufficient number of empty entries to encode~$\mathbf{I}$, we get the following restriction on $\mathbf{d}$:
    \[
    n^{d}-\ell_1 \cdot \ell_2 \cdot \ldots \cdot \ell_{d} +\left\lceil d\cdot \log(n)\right\rceil \leq n^{d}-1,
    \]
    and hence, 
    $\left\lceil d\cdot \log(n)\right\rceil + 1 \leq \ell_1 \cdot \ell_2 \cdot \ldots \cdot \ell_{d}$. In the special case where $\ell(n) = \ell_1=\ell_2=\ldots =\ell_{d}$ the latter bound is equivalent to
    \[
    \left\lceil\sqrt[d]{\left\lceil d\cdot \log(n)\right\rceil + 1}\right\rceil \leq \ell(n).
    \]
    \end{construction}
    The supported values of $\ell(n)$ based on our construction match the construction presented in~\cite{marcovich2023zero}, which requires a more sophisticated convergence-based proof. For the binary alphabet case, this result also coincides with the lower bound for $\ell(n)$ derived in \cite{marcovich2023zero} using a union bound argument, under the condition that at most a single redundancy bit is used.

    The authors of~\cite{marcovich2023zero} also extended the ZRCF constraint to support rectangular cuboids of arbitrary size but with a constant $d$-volume $V$.
\begin{definition}
    ($d$-Volume). Given an array $A$ of size $\mathbf{d}=(\ell_1,\ell_2,\ldots,\ell_{d})$, the  $d$-volume of $A$, denoted by $Vol(A)$, is defined to be $\ell_1\cdot \ell_2\cdot\ldots\cdot \ell_{d}$. 
\end{definition}
Given an array $A\in\Sigma^{n^d}$, this generalization prohibits any sub-array of zeros with $d$-volume $V\triangleq V(n,d)$, i.e.,
\begin{align*}
        \mathcal{C}_{V-ZRCF}\left(n, d\right) =  \left\{ A\in\Sigma^{n^{d}}: \hspace{-2ex}
        \begin{array}{c}
    \forall \mathbf{I},\mathbf{d}, \text{ if }  \\ Vol(A_{\mathbf{I},\mathbf{d}}) \ge V,   \\ \text{ then  } A_{\mathbf{I},\mathbf{d}} \ne \overline{\mathbf{0_d}}\end{array} \right\} .
\end{align*}
To address this constraint, the authors estimated the number of minimal arrays for a given \( d \)-volume $V$, denoted by \( f_d(V) \), and showed that \( f_d(V) = \Theta(V^{\frac{d-1}{d}}) \). Using this, they presented a pair of encoder and decoder for the $\mathcal{C}_{V-ZRCF}(n,d)$ constraint that uses  \( \log(f_d(V)) \) bits to represent the size $\mathbf{d}$ of an all-zero deleted sub-array (see \cite[Algorithm 5]{marcovich2023zero}). 

By modifying the mapping $\xi$ of Construction~\ref{const: ZRCF} to use this mapping in addition to the starting location $\mathbf{I}$, we found that the universal approach can support the same values of $V$ as in \cite{marcovich2023zero}. Namely, 
\begin{align*}
\left\lceil d \cdot \log(n) \right\rceil \hspace{-0.5ex}+ \hspace{-0.5ex}\left\lceil \frac{d-1}{d} \cdot \log(\log(n)) \right\rceil + C + 1 \leq V(n),
\end{align*}
where \( C \) is a constant that depends on \( d \), with its explicit form provided in \cite{marcovich2023zero}. 
We note that the implementation of this mapping was not given in the paper, and its existence follows from counting arguments.

\subsection{Repeat-Free}
The $\mathbf{d}$-Repeat-Free ($\mathbf{d}$-RF) constraint prohibits two or more identical sub-arrays of size $\mathbf{d}$. That is,
\[
\mathcal{C}_{\mathbf{d}-RF}\left(n, d\right)=\left\{ A\in\Sigma^{n^{d}}:\begin{array}{c}
\text{for all }\mathbf{I_1}\neq\mathbf{I_2}\\
A_{ \mathbf{I_1}, \mathbf{d}}\ne A_{\mathbf{I_2},\mathbf{d}}
\end{array}\right\} 
\]
Similarly to the previous constructions, our approach is based on deleting problematic sub-arrays. Clearly, if we have two identical and non-overlapping sub-arrays and we delete only one of them, by knowing their starting locations $\mathbf{I_1}$ and $\mathbf{I_2}$, one can be recovered from the other. Next, we show that even when two identical sub-arrays overlap, it is still possible to reverse the process and recover the overlapping part (and the rest of the deleted sub-array).
\begin{lemma}\label{lem: rf}
For any  $n, d, \mathbf{d}$ and $A\notin \mathcal{C}_{\mathbf{d}-RF}\left(n, d\right)$ such that $A_{ \mathbf{I_1}, \mathbf{d}} = A_{\mathbf{I_2},\mathbf{d}}$ are two overlapping sub-arrays with starting indices $\mathbf{I_1}\ne \mathbf{I_2}$, $A$ can be recovered from $\mathbf{I_1}, \mathbf{I_2}$ and $R(A_{\mathbf{I_2}, \mathbf{d}})$.
\end{lemma}
\begin{IEEEproof}
Due to space limitations, we only present the proof idea for $d=2$, while the proof for $d>2$ can be done using similar arguments (and the proof for $d=1$ can be found in~\cite{bar2023universal}). 

As $\mathbf{I_1}, \mathbf{I_2}$ are known, we can reverse the sub-array deletion process such that the empty entries are at the correct location, i.e., the location that corresponds with the deleted sub-array $A_{\mathbf{I_2}, \mathbf{d}}$. 
Assume w.l.o.g. that  $\mathbf{I_1}=(x_1,y_1)$ and $\mathbf{I_2}=(x_2,y_2)$ differ in the first coordinate and further assume that $x_1<x_2$. As $x_1<x_2$, all the entries of the form $(x_1+x,y_1+y)$ of $A_{ \mathbf{I_1}, \mathbf{d}}$, such that $x\le x_2-x_1$ and $y=0,1,\ldots, \ell_2$ are not part of the overlap. Furthermore, as $A_{ \mathbf{I_1}, \mathbf{d}} = A_{\mathbf{I_2},\mathbf{d}}$, we know that these are exactly the values of the first rows of $A_{\mathbf{I_2},\mathbf{d}}$, i.e., $(x_1+x,y_1+y) = (x_2+x,y_2+y)$ for $x\le x_2-x_1$ and $y=0,1,\ldots, \ell_2$. As $x_2>x_1$, by doing so we filled at least  one additional row of  $A_{ \mathbf{I_1}, \mathbf{d}}$ that was incomplete as it was part of the overlap, which allows as to continue with the process for larger values of $x$ until all the entries are recovered. This process is illustrated in  Figure~\ref{fig: sub-array rf}.  
    \begin{figure}[h]
    \centering \includegraphics[width=.95\linewidth]{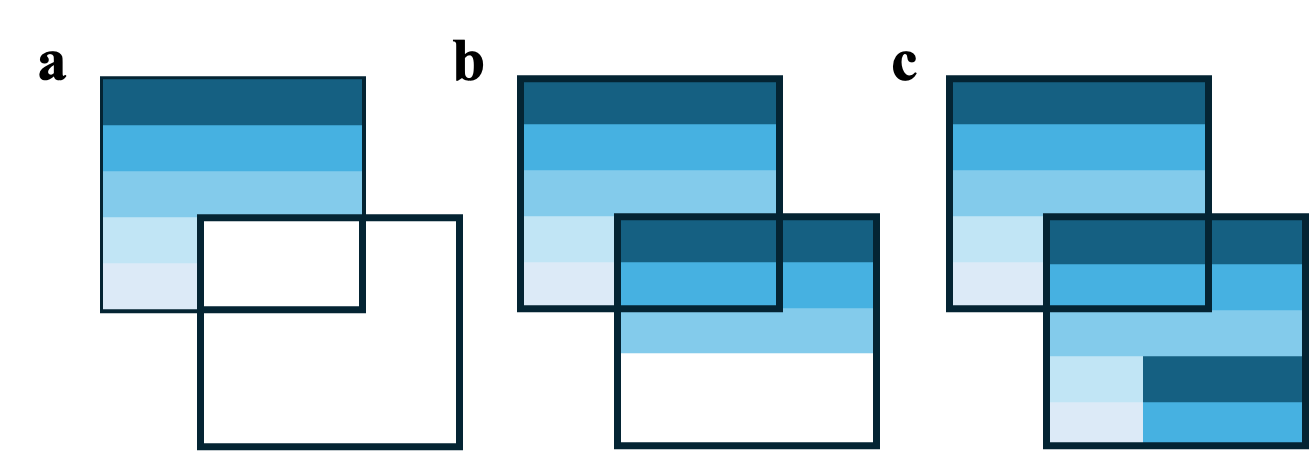}
    \caption{Illustration of the process described in Lemma~\ref{lem: rf}  }
    \label{fig: sub-array rf}
\end{figure}
\end{IEEEproof}

\begin{construction}
Define $\mathbf{1}_{\mathcal{C}(n,d)}:\Sigma^{n^d}\to\Sigma$ to be a mapping that scans all the pairs of sub-arrays of size $\mathbf{d}$ and returns $1$ if and only if there is no pair of identical sub-arrays of size $\mathbf{d}$. Additionally, define 
 $\xi:\overline{
    \mathcal{C}_{\mathbf{d}-RF}\left(n, d\right)}\rightarrow\Sigma^{n^{d}-1}$
    by 
    \[
\xi\left(A\right)=R\left(A_{\mathbf{I_2},\mathbf{d}}\right)\circ \mathbf{I_1}\circ \mathbf{I_2},
\]
where $\mathbf{I_1}$ is the first location for which the constraint is violated and $\mathbf{I_2}$ is the first location such that $A_{ \mathbf{I_1}, \mathbf{d}} = A_{\mathbf{I_2},\mathbf{d}}$.
As $A\in\overline{C_{\mathbf{d}-RF}\left(n, d\right)}$ such $\mathbf{I}_1$ and $\mathbf{I}_2$ exist and by  Lemma~\ref{lem: rf} this function is injective as required by Theorem ~\ref{th: construction existence}.
To have a sufficient number of empty entries, we have that
\[
n^{d}-\ell_1 \cdot \ell_2 \cdot \ldots \cdot \ell_{d} + 2\cdot\left\lceil d\cdot \log(n)\right\rceil \leq n^{d}-1,
\]
and hence,
\[
 2\cdot \left\lceil d\cdot \log(n)\right\rceil + 1 \leq \ell_1 \cdot \ell_2 \cdot \ldots \cdot \ell_{d}.
\]
In the special case where
$\ell(n) = \ell_1=\ldots =\ell_{d}$ we obtain
\[
\left\lceil\sqrt[d]{ 2\cdot \left\lceil d\cdot \log(n)\right\rceil + 1}\right\rceil \leq \ell(n).
\]
\end{construction}

For \( d = 2 \), a construction using a single redundancy bit is given in~\cite{marcovich2023zero}. While the convergence of the algorithm in~\cite{marcovich2023zero} is shown using a sophisticated proof, the construction only supports  \( \ell(n) \geq 2 \cdot \left\lceil \sqrt{\left\lceil 3 \cdot \log(n) \right\rceil + 2} \right\rceil \) which further highlights the strength of our suggested universal construction.  
We also note that using a union bound argument, a lower bound on \( \ell(n) \) for which there exists a single redundancy bit construction is presented in~\cite{marcovich2023zero}. More specifically, for a binary alphabet, it was shown that the lower bound on the achievable $\ell(n)$ is
$\sqrt[d]{2 \cdot d \cdot \log(n) + 1}$
which matches our construction's result up to the ceiling functions.

\subsection{Hamming Distance Repeat-Free}
An interesting generalization of the Repeat-Free constraint might be a constraint that forbids two sub-arrays from being too similar to each other. More formally, for two arrays of the same size $A,B$, the Hamming distance between $A$ and $B$, $d_H(A,B)$, is the number of symbols on which $A$ and $B$ do not agree. The $(\mathbf{d},p)$-Hamming Distance Repeat-Free ($(\mathbf{d},p)$-HDRF) constraint requires that any two sub-arrays of size $\mathbf{d}$ are of Hamming distance at least $p=p(n)$, i.e.,

\begin{align*}
\mathcal{C}_{(\mathbf{d},p)-HDRF}\left(n, d\right)\hspace{-.3ex}=\hspace{-.3ex}
\left\{ \hspace{-0.6ex}A\in\Sigma^{n^{d}}\hspace{-1.5ex}:\hspace{-1.5ex}\begin{array}{c}
\text{for all }\mathbf{I_1}\neq\mathbf{I_2}\\
 d_H(A_{\mathbf{d}, \mathbf{I_1}},A_{\mathbf{d}, \mathbf{I_2}})\ge p 
\end{array}\hspace{-1.3ex}\right\} 
\end{align*}

\begin{construction}
Define $\mathbf{1}_{\mathcal{C}(n,d)}:\Sigma^{n^d}\to\Sigma$ to be a mapping that scans all the pairs of sub-arrays of size $\mathbf{d}$ and returns $1$ if and only if there is no pair of sub-arrays of size $\mathbf{d}$ and Hamming distance smaller than $p$. Additionally, define 
 $\xi:\overline{\mathcal{C}_{(\mathbf{d},p)-HDRF}\left(n, d\right)}
    \rightarrow\Sigma^{n^{d}-1}$
    for $A\in\overline{\mathcal{C}_{(\mathbf{d},p)-HDRF}}$ by 
\[
\xi\left(A\right)=R\left(A_{\mathbf{I_2},\mathbf{d}}\right)\circ \mathbf{I_1}\circ \mathbf{I_2}\circ \mathbf{P_1}\circ \mathbf{P_2}\circ \cdots\circ \mathbf{P_{p(n)-1}},
\]
where $\mathbf{I_1}$ is the first location for which the constraint is violated, $\mathbf{I_2}$ is the first location such that ${d_H(A_{ \mathbf{I_1}, \mathbf{d}},  A_{\mathbf{I_2},\mathbf{d}})< p}$,
and $\mathbf{P}_j$ are the 
$\lceil\log(\ell_1\cdot \ldots\cdot \ell_d + 1)\rceil$ 
bits encoding of the locations in which $A_{ \mathbf{I_1}, \mathbf{d}}$ and $  A_{\mathbf{I_2},\mathbf{d}}$ differ\footnote{To extend this construction to the non-binary case the symbols in these locations must also be encoded.}.  
Note that $\xi$ is injective by arguments similar to those presented in Lemma~\ref{lem: rf} and as we reserved a sufficient number of bits for the encoding of the locations $\mathbf{P}_j$, we can use a dummy location if the Hamming distance is strictly less than $p-1$. To have a sufficient number of empty entries, we require that
\begin{align*}
    n^d& -(\ell_1\cdot\ldots\cdot\ell_d)  +\left\lceil 2\cdot d\cdot \log(n)\right\rceil\\ &+ (p-1)\lceil\log(\ell_1\cdot \ldots\cdot \ell_d + 1)\rceil \le n^d-1,
\end{align*}
which is equivalent to 
\begin{align*}
        \left\lceil 2\cdot d\cdot \log(n)\right\rceil& + (p-1)\lceil\log(\ell_1\cdot \ldots\cdot \ell_d + 1)\rceil+1 \\ & \le \ell_1\cdot\ldots\cdot\ell_d.
\end{align*}
 
In the special case where
$\ell(n) = \ell_1=\ldots =\ell_{d}$ we obtain that
\begin{align*}
        \sqrt[d]{\left\lceil 2\cdot d\cdot \log(n)\right\rceil + (p-1)\lceil\log(\ell(n)^d + 1)\rceil+1}  \le \ell(n).
\end{align*}
\end{construction}

\section{Discussion}
This paper presents a universal framework to address multi-dimensional parametric constraints, where the conditions on allowed arrays are functions of their size. Specifically, the framework is demonstrated on three distinct parametric constraints:
\begin{enumerate}
    \item The $\mathbf{d}-ZRCF$ constraint:
     In this case, a valid array cannot contain all-zero sub-arrays of size $\mathbf{d}$. Our algorithm achieves the same results as the state-of-the-art method from~\cite{marcovich2023zero} and meets the best known lower bound established in~\cite{marcovich2023zero}.
    \item The $\mathbf{d}-RF$ constraint:
     Here, no two identical sub-arrays of size $\mathbf{d}$ are allowed. Using our universal approach, we improve the construction presented in~\cite{marcovich2023zero}, providing an explicit construction that matches the lower bound derived in~\cite{marcovich2023zero} (up to the ceiling function).
    \item The $(\mathbf{d},p)$-HDRF constraint:
    This previously unsolved constraint requires that any two sub-arrays of size $\mathbf{d}$ be at least $p$ Hamming distance apart.
\end{enumerate}
In Section~\ref{sec:background}, we propose a simple method for deleting sub-arrays. While this method was chosen for its simplicity, a more efficient alternative would be to delete the sub-array directly from the array $A$ and then swap entries to propagate the empty entries to the desired locations.

While this work represents a significant step in advancing the construction of multi-dimensional parametric constrained codes, there remains considerable room for improvement and several open problems to address in future research. Notably, we suggest the following directions:
\begin{enumerate}
    \item \textbf{Time Complexity Analysis}: A thorough analysis of the time complexity for the proposed construction, as well as other constructions derived from the universal framework, is necessary for assessing scalability and efficiency
    \item \textbf{Handling More Complex Constraints}: The current framework primarily addresses constraints that involve the deletion of sub-arrays. Future work should aim to handle more \emph{global} constraints, where no specific sub-array can be identified as problematic. One example is the almost-balanced constraint studied in the one-dimensional setup by~\cite{bar2024balance}. While the approach in~\cite{bar2024balance} can be extended to the $d$-dimensional case to balance the entire array, challenges remain, e.g., balancing one-dimensional rows along all axes of the array.
    \item \textbf{Development of Lower Bounds}: Future research should focus on developing lower bounds for the achievable parameters under different constraints and evaluating how the universal framework performs relative to these bounds.
    \item \textbf{Extension to Multiple Redundancy Symbols}: The framework should be extended to support more than one redundancy symbol, particularly for cases where the parameters of interest cannot support constructions with a single redundancy symbol.
\end{enumerate}

\section*{Acknowledgment}
The authors sincerely thank Eyal Barak for his contributions to the project that served as the foundation for this work and Ronny M. Roth for insightful discussions and helpful suggestions.

\newpage
\bibliographystyle{IEEEtran}
\bibliography{refs}

\balance

\end{document}